\documentclass[useAMS,usenatbib]{mn2e}

\usepackage{graphicx}
\usepackage{amsmath}
\usepackage{amssymb}

\title[Heating Rate Profiles in Galaxy Clusters]{Heating Rate Profiles
in Galaxy Clusters}

\author[E.C.D. Pope, G. Pavlovski, C.R. Kaiser, H. Fangohr] {Edward
C.D. Pope$^{1,2}$\thanks{E-mail:edpope@soton.ac.uk}, Georgi
Pavlovski$^{1}$, Christian R. Kaiser$^{1}$, Hans Fangohr$^{2}$ \\
$^{1}$School of Physics \& Astronomy, University of Southampton, UK,
SO17 1BJ\\ $^{2}$School of Engineering Sciences, University of
Southampton, UK, SO17 1BJ}

\begin{document}

\pagerange{\pageref{firstpage}--\pageref{lastpage} \pubyear{2005}}

 \maketitle

\label{firstpage}

\begin{abstract}

In recent years evidence has accumulated suggesting that the gas in
galaxy clusters is heated by non-gravitational processes. Here we
calculate the heating rates required to maintain a physically motived
mass flow rate, in a sample of seven galaxy clusters. We employ the
spectroscopic mass deposition rates as an observational input along
with temperature and density data for each cluster. On energetic
grounds we find that thermal conduction could provide the necessary
heating for A2199, Perseus, A1795 and A478. However, the suppression
factor, of the clasical Spitzer value, is a different function of
radius for each cluster. Based on the observations of plasma bubbles
we also calculate the duty cycles for each AGN, in the absence of
thermal conduction, which can provide the required energy input. With
the exception of Hydra-A it appears that each of the other AGNs in our
sample require duty cycles of roughly $10^{6}-10^{7}$ yrs to provide
their steady-state heating requirements. If these duty cycles are
unrealistic, this may imply that many galaxy clusters must be heated
by very powerful Hydra-A type events interspersed between more
frequent smaller-scale outbursts. The suppression factors for the
thermal conductivity required for combined heating by AGN and thermal
conduction are generally acceptable. However, these suppression
factors still require `fine-tuning` of the thermal conductivity as a
function of radius. As a consequence of this work we present the AGN
duty cycle as a cooling flow diagnostic.

\end{abstract}


\section{Introduction}

The radiative cooling times of the hot, X-ray emitting gas in the
centres of many galaxy clusters may be as short as $10^{6}$ years. In
the classical model \citep[see e.g.][for a review]{fab94}, where
radiative losses occur unopposed, a cooling flow develops in which
the gas cools below X-ray temperatures and accretes onto the central
cluster galaxy where it accumulates in molecular clouds and
subsequently forms stars. However, studies have demonstrated both a
lack of the expected cold gas at temperatures well below 1-2keV
\citep[e.g.][]{edge01} and that spectoscopically determined mass
deposition rates are a factor of 5-10 less than the classically
determined values \citep{voigt04}. These findings suggest that the
cluster gas is being reheated in order to produce the observed minimum
temperatures and the reduction in excess star formation compared to
expectations. As yet it is unclear whether this heating occurs
continuously or periodically.

There is observational evidence that the temperature profiles of
galaxy cluster atmospheres are well described by the same mathematical
function across a range in redshift \citep[e.g.][]{allen01}. Given
such similarity it is possible that the atmospheres of galaxy clusters
are in a quasi-steady state and that observable parameters such as the
radial temperature profile do not vary signicantly over the lifetime
of a cluster. This universal temperature profile would be difficult to
sustain if the density profiles varied significantly with time. In
addition, if large quantities of gas were deposited in the central
regions one might expect that the density profiles would be much more
centrally peaked than observations suggest. The culmination of this
argument is that, at least in the central regions, the inward flow
rate of mass must be roughly independent of radius to ensure that mass
is not deposited at any particular location which would significantly
alter the density profile. 

One particular problem with this is that all of the mass flowing in
this region must be deposited at the cluster centre. Therefore, unless
the mass flow rate is small this would result in large quantities of
gas at the cluster centre.

In contrast, spectroscopically determined mass deposition rates
suggests that mass $\it{is}$ deposited at a roughly constant rate
throughout the cluster, at resolved radii
\citep[][]{voigt04}. However, a roughly constant mass deposition rate
is indicative of a mass flow rate that is roughly proportional to
radius, rather than constant as the density profiles suggest. Yet, if
such a flow were persistent this proportionality would result in
density profiles which contradict the observational evidence by being
more centrally peaked.

If we are to reconcile the implications of the density profiles and
mass deposition rates then one possible explanation is that the mass
flow rate exhibits two different asymptotic properties. That is, near
the cluster centre or within the central galaxy the mass flow rate
should be roughly constant so that the density profiles are
essentially left unchanged, while the mass deposited at the cluster
centre is sufficiently small to agree with observations. At larger
radii the mass flow rate should still be proportional to radius and
satisfy the observations on resolvable scales which imply a constant
mass deposition rate. The relationship between the mass flow rates and
the mass deposition rates are defined section 3.

The two main candidates for heating cluster atmospheres are: Active
Galactic Nuclei (AGN) \citep[e.g.][]{tabor93,bub01,nature,brueggen03}
and thermal conduction \citep[e.g.][]{gaetz, zakamska03, voigt02,
voigt04}. Heating by AGN is thought to occur through the dissipation
of the internal energy of plasma bubbles inflated by the AGN at the
centre of the cooling flow. Since these bubbles are less dense than
the ambient gas, they are buoyant and rise through the intracluster
medium (ICM) stirring and exciting sound waves in the surounding gas
\citep[e.g.][]{ruszbegel02,fabpers02}. This energy may be dissipated
by means of a turbulent cascade, or viscous processes if they are
significant. Deep in the central galaxy other processes such as
supernovae and stellar winds will also have some impact on the ambient
gas.

However, AGN are only periodically active which could result in
similarly periodic heating rates, thus making the possibility of a
totally steady state unlikely. In this case one could imagine a
scenario in which a quasi steady-state is possible where temperature
and density profiles oscillate around their average values. It is also
possible that dissipation of the plasma bubbles may occur over
timescales longer than the AGN duty cycle thus providing almost
continous heating \citep[e.g.][]{reynolds05}.

Thermal conduction may also play a significant role in transferring
energy towards central regions of galaxy clusters given the large
temperature gradients which are observed in many clusters. In fact,
several authors \citep[e.g.][]{voigt02,voigt04} have shown that on
energy grounds alone it may be possible for thermal conduction to
provide the necessary heating in some clusters.

However, the exact value of the thermal conductivity remains
uncertain. The theoretical value for the thermal conductivity of a
fully ionised, unmagnetised plasma is calculated by
\cite{spitzer}. Magnetic fields, which are thought to exist in the
cluster gas, based on Faraday rotation measures of clusters
\citep[e.g.][]{carilli02}, may greatly alter this value. The standard
method by which the unknown effects of magnetic fields are taken into
account is to include a suppression factor, $f$, in the Spitzer
formula, which indicates the actual value as a fraction of the full
Spitzer value.

Several authors have found temperature profiles in some galaxy
clusters that are compatible with thermal conduction, and values of
the suppression factor that are physically meaningful
\citep[e.g.][]{zakamska03}. In contrast, more detailed work in which
the Virgo cluster was simulated using complete hydrodynamics,
including radiative cooling and heating by thermal conduction has
demonstrated that thermal conduction cannot prevent the occurence of a
cooling catastrophe, in this particular example
\citep[][]{pope05}. Furthermore, the mass flow rates under such
circumstances are not constant with radius so that mass builds up in
particular areas and therefore changes the density distribution,
making it much more centrally peaked than observed in real clusters.

To answer the questions regarding the necessary heating rates and
mechanisms involved in galaxy clusters we construct a simple model
based on two main assumptions: galaxy clusters are in approximate
steady state, and the mass flow rates fulfill the constraints provided
by observations at both large and small radii. To ensure this, the
radial behaviour of the mass flow rates is modelled using an
observationally motivated, but empirical function consistent with the
observations. Using mathematical functions fitted to the temperature
and density data of seven galaxy clusters we derive the required
heating rates within regions which are consistent with those for which
the mass flow rates were determined. We also compare the observed
energy currently available in the form of plasma bubbles with the
heating requirements for each cluster. From the heating rate profiles
we calculate the thermal conduction suppression factor as a radial
function for each cluster in order that we may determine which
clusters could be heated by this process.

The plan for this paper is as follows. In Section 2 we give the
details of the parameters used to fit observational temperature and
density data for seven galaxy clusters. In Section 3 and 4 we describe
the model used to estimate the required heating, and thermal
conduction suppression factors. The heating rates we have calculated
are compared with observational estimates of both AGN heating and
a combination of AGN heating and thermal conduction in Sections 5 and
6. In Section 7 we summarize our main findings.

The results are given for a cosmology with $H_{0}=70{}{\rm km
s}^{-1}{\rm Mpc}^{-1}$, $\Omega_{M} = 0.3$ and $\Omega_{\Lambda} =
0.7$.

\section{Functions fitted to observational data}

\begin{figure*}
\centering \includegraphics[width=14cm]{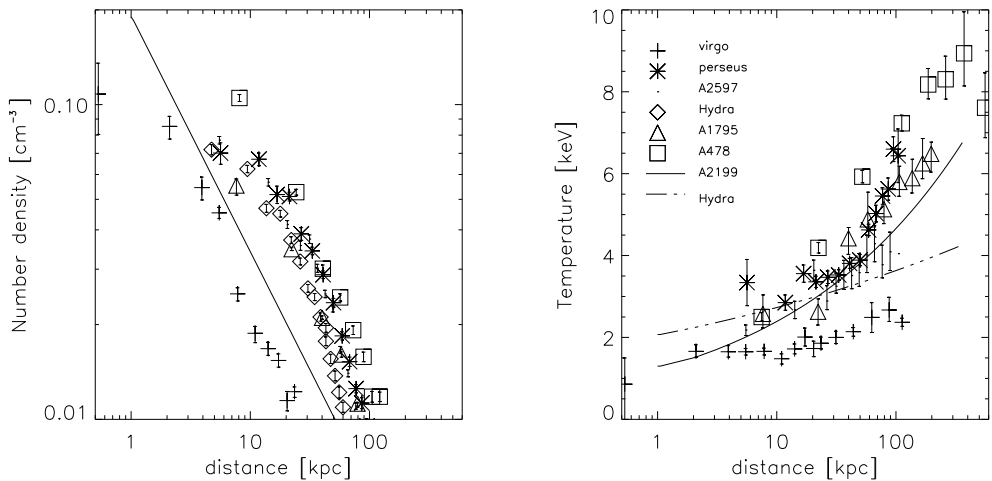}
\caption{Left: density data for all clusters except A2199 for which we
show the best-fit, since we did not have access to the observational
data. Right: temperature data except for A2199 and Hydra for which we
give the fitted functions, since we did not have access to the
observational data. The key in the right-hand panel applies to both
temperature and density data.}
\label{fig:fitdens}
\end{figure*}

In this section we give details of the functions used to derive the
mass deposition and heating rate profiles, the references to the
observational data, and show a comparison with the data in figure
\ref{fig:fitdens}.

\begin{table*}
\centering
\begin{minipage}{140mm}
\begin{tabular}{rcccccc}
\hline Name & $n_{0}({\rm cm}^{-3})$ & $n_{1}({\rm cm}^{-3})$& $\beta_{0}$&
$\beta_{1}$& $r_{0}({\rm kpc})$& $r_{1}({\rm kpc})$ \\ \hline

Virgo & $0.089{\pm 0.011}$ & $0.019 \pm{0.002}$ & $1.52 {\pm 0.32}$ &
$0.705{}$& $5{\pm 1}$ & $23.3{\pm 4.3}$\\
\\
Perseus & $0.071{\pm 0.003}$ & $$ & $0.81 {\pm 0.04}$ &
$$& $28.5{\pm 2.7}$ & $$\\
\\
Hydra & $0.07{\pm 0.02}$ & $$ & $0.72 {\pm 0.004}$ &
$$& $18.6{\pm 0.5}$ & $$\\
\\
A2597 &  $0.13{\pm 0.07}$ & $$ & $0.15 {\pm 0.10}$ &
$0.79{\pm 0.06}$& $1.0{\pm 2.6}$ & $43{\pm 14}$\\
\\
A2199 & $0.19{}$ & $$ & $0.75 {}$ &
$$& $1$ & $$\\
\\
A1795 & $0.066{\pm 0.067}$ & $$ & $0.24 {\pm 0.11}$ &
$0.41{\pm 0.13}$& $57{\pm 37}$ & $12{\pm 4.9}$\\
\\

A478 & $0.153{\pm 0.019}$ & $$ & $0.55 {\pm 0.05}$ &
$0.41{\pm 0.34}$& $145{\pm 32}$ & $6.6{\pm 1.9}$\\

\hline
\end{tabular}
\caption{Summary of best fit parameters for density profiles with
1-sigma errors obtained from the least-squares fitting procedure. We
do not give errors the model function parameters for A2199 since we
were unable to obtain the observational data. See table 2 for
references.}\label{tab:1}
\end{minipage}
\end{table*}

We have only fitted our own functions to the data when the original
authors had not included confidence intervals for their model
parameters, but we give the functions used to fit all of the cluster
atmospheres. The values of the fitted parameters are given in tables
\ref{tab:1} and \ref{tab:2}.

The Virgo cluster`s electron number density data was fitted by
\cite{ghizzardi04} using a double $\beta$-profile,

\begin{equation} \label{eq:vdens}
n = \frac{n_{0}}{[1+(r/r_{0})^{2}]^{\beta_{0}}} +
\frac{n_{1}}{[1+(r/r_{1})^{2}]^{\beta_{1}}}.
\end{equation}

For simplicity, the Perseus \citep[see][]{sanders} and Hydra
\citep[][]{davhyd01} density data were fitted with single
$\beta$-profiles,

\begin{equation} \label{eq:phdens}
n = \frac{n_{0}}{[1+(r/r_{0})^{2}]^{\beta_{0}}}.
\end{equation}

We were unable to obtain the observational data for A2199 and so
relied on the fits presented by the original authors
\citep[][]{johnstone} who found that a simple power-law was sufficient
to describe the density distribution,
\begin{equation} \label{eq:A2199dens}
n = n_{0}\bigg(\frac{r}{r_{0}}\bigg)^{-\beta_{0}}.
\end{equation}

For A2597, A1795 and A478
\citep[see][respectively]{mcnam01,ettori,sun} we employ the function
used by \cite{denchand} to fit the density distributions,
\begin{equation} \label{eq:Asdens}
n=\frac{n_{0}}{[1+(r/r_{0})^{2}]^{\beta_{0}}}\frac{1}{[1+(r/r_{1})^{2}]^{\beta_{1}}}.
\end{equation}

\begin{table*} 
\centering
\begin{minipage}{140mm}
\begin{tabular}{rccccc}
\hline Name & $T_{0} ({\rm keV})$ & $T_{1}({\rm keV})$& $r_{ct}({\rm
kpc})$& $\delta$ & Reference\\ 
\hline 
Virgo & $2.4{\pm 0.1}$ &$0.77{\pm 0.10}$ & $23{\pm 4}$ & $$ & $1$\\ 
\\ 
Perseus & $8.7{\pm 1.1}$ & $$ & $63{\pm 5}$ & $2.7{\pm 0.3}$ & $2$\\ 
\\ 
Hydra &$2.73{\pm 0.07}$ & $$ & $10{ }$ & $0.12{\pm 0.01}$ & $3$\\ 
\\
A2597 & $4.1{\pm 0.1}$ & $2.0{\pm 0.2}$ & $16{\pm 10}$ &$0.7{\pm 0.4}$ & $4,5$\\ 
\\ 
A2199 & $1.0{}$ & $$ & $1$ & $0.29$ &$6$\\ 
\\ 
A1795 & $7.2{\pm 1.9}$ & $5.1{\pm 2.2}$ & $40{\pm 34}$& $0.5{\pm 0.7}$ & $7,5$ \\ 
\\ 
A478 & $9.7{\pm 1.5}$ & $7.6{\pm 1.7}$ & $16{\pm 5}$ & $0.3{\pm 0.2}$ & $8,5$\\ 
\hline
\end{tabular}
\caption{Summary of temperature parameters fitted to the data with
1-sigma errors obtained from the least-squares fitting procedure.
References:-(1) Ghizzardi et al. (2004); (2) Sanders et al. (2004);
(3) David et al. (2001); (4) McNamara et al. (2001) and Rafferty et
al. (2005), in preparation; (5) Dennis \& Chandran (2005); (6)
Johnstone et al. (2002); (7) Ettori et al. (2002); (8) Sun et
al. (2003). We do not give errors for A2199 since we were unable to
obtain the observational data. We also do not give an error for the
Hydra temperature scale-height since none was given in the original
publication, David et al. (2001). }\label{tab:2}
\end{minipage}
\end{table*}

As above, for the cluster gas temperature profiles we have only fitted
functions to the data when the original authors had not included
confidence intervals, but give the details for all clusters in our
sample.

\cite{ghizzardi04} found that the Virgo temperature data is best
described by a Gaussian curve,
\begin{equation} \label{eq:vtemp}
T = T_{0} -T_{1}\exp\bigg(-\frac{1}{2}\frac{r^{2}}{{r_{\rm ct}}^{2}}\bigg).
\end{equation}

\cite{churpers} fitted the Perseus temperature data with the function,

\begin{equation} \label{eq:ptemp}
T = T_{0} \bigg[\frac{1 + \big(r/r_{\rm ct}\big)^{3}}{\delta
+\big(r/r_{\rm ct}\big)^{3}}\bigg].
\end{equation}

We fitted the temperature data for A2597, A1795 and A478
using the same function as \cite{denchand},
\begin{equation} \label{eq:PAstemp}
T = T_{0} -\frac{T_{1}}{\big[1 + {(r/r_{\rm ct})}^{2}\big]^{\delta}}.
\end{equation}

The best-fit temperature profile for the Hydra cluster is given by
\cite{davhyd01} as a power-law,
\begin{equation} \label{eq:Htemp}
T = T_{0} \bigg(\frac{r}{r_{\rm ct}}\bigg)^{\delta}.
\end{equation}

The A2199 temperature data was found by the original authors
\citep[][]{johnstone} to be described by a power-law without a
characteristic length-scale. For consistency we use equation
(\ref{eq:Htemp}) to represent the temperature distribution in this
cluster.

Although we have provided 1-sigma errors for the fitted parameters
that describe the temperature and density distribution of the gas in
galaxy clusters, we will not give errors for any quantities derived
from these parameters. The reason for this is that the errors and
gradients of the temperature and density are all model dependent. In
addition, the calculations are complex, making the standard
propagation of errors impractical and error estimates themselves very
uncertain. In addition, although we can make very conservative
estimates of the confidence limits, of a particular derived quantity,
they are too vague to be of importance.

\section{The Model}

\subsection{General heating rates}

Starting from the assumption that the atmospheres of galaxy clusters
are spherically symmetric, and in a quasi steady-state, it is possible
to derive what the time averaged heating rate, as a function of
radius, must be in order to maintain the observed temperature and
density profiles,

\begin{equation} \label{eq:h0}
h = n^{2}\Lambda_{\rm rad} + \frac{1}{r^{2}}\frac{\rm d}{{\rm d}
r}\bigg[\frac{\dot{M}}{4 \pi}\bigg(\frac{5k_{\rm b}T}{2\mu m_{\rm p}}
+ \phi \bigg)\bigg],
\end{equation}
where $\dot{M}$ is the mass flow rate, $T$ is the gas temperature, $n$
is the gas number density, $\Lambda_{\rm rad}$ is the cooling function
and $\phi$ is the gravitational potential. Here we assume only
subsonic flow of gas.

The mass flow rate through a spherical surface, radius $r$, is
\begin{equation} 
\dot{M} = 4\pi r^{2} \rho v_{\rm r},
\end{equation}
where $\rho$ is the gas density and $v_{\rm r}$ is the gas velocity in
the radial direction.

The difference between the rate at which mass enters and leaves a
spherical shell, of thickness $\Delta r $, is called the mass
deposition rate, $\Delta \dot{M}$. In terms of the mass flow rate, the
mass deposition rate is
\begin{equation} 
\Delta \dot{M} \approx \bigg(\frac{{\rm d}\dot{M}}{{\rm d}
r}\bigg)\Delta r
\end{equation}

Consequently, the mass flow rate, at a particular radius, can be
recovered by summing the contributions of the mass deposition rate in
each shell up to that radius. Henceforth, we refer to the mass flow
rate as the integrated mass deposition rate.

For subsonic gas flow the cluster atmosphere can be assumed to be in
approximate hydrostatic equilibrium. On this basis, we are able to
estimate the gravitational acceleration from the temperature and
density profiles of the cluster atmosphere,

\begin{equation} \label{eq:hydro}
\frac{{\rm d} P}{\rm d r} = -\rho\frac{{\rm d} \phi}{\rm d r},
\end{equation}
where $P$ is the gas pressure and $\rho$ is the gas density.

The gas pressure is related to density and temperature by the ideal
gas equation

\begin{equation} \label{eq:pressure}
P = \frac{\rho k_{\rm b}T}{\mu m_{\rm p}},
\end{equation}
where $\mu m_{\rm p}$ is the mean mass per particle. We assume that
$\mu$=16/27 as appropriate for standard primordial abundance.

To avoid anomalies when calculating spatial derivatives, we fit
continuous analytical functions through the density and temperature
data in the previous section. This ensures that we do not encounter
any large discontinuities which may result in extreme heating rates.

The only unknowns in equation (\ref{eq:h0}) are then the integrated
mass deposition rate, $\dot{M}$, and the heating rate, $h$.

\subsection{Calculating Mass Deposition rates}

In the classical cooling flow picture, which assumes that the heating
is zero everywhere, and ignoring the effect of the gravitational
potential, the bolometric X-ray luminosity, integrated mass deposition
rate and gas temperature are related by integrating equation
(\ref{eq:h0}) \citep[e.g.][]{fab94} over a spherical
volume. Analytically, the bolometric X-ray luminosity within a given
radius is simply the volume integral of the radiative cooling rate per
unit volume,

\begin{equation} \label{eq:Lx}
L_{\rm X}(<r) = 4 \pi \int_{0}^{r} n^{2}\Lambda_{\rm rad}r^{2}{{\rm d}
r}.
\end{equation}

Integrating the second term in equation (\ref{eq:h0}) gives the
reationship between bolometric luminosity, integrated classical mass
deposition rate and temperature,

\begin{equation} \label{eq:Lx1}
L_{\rm X}(<r) = \frac{5k_{\rm b}}{2 \mu m_{\rm p}}\dot{M}_{\rm
clas}(r)T(r).
\end{equation}

For observational data the luminosity and classical mass deposition
rates at each radius are estimated by dividing the observed 2-d
projection of the cluster into concentric shells. Using this method
one is able to reconstruct the 3-d properties of a cluster. The
contribution, to the total X-ray luminosity, of the $jth$ spherical
shell, of thickness $\Delta r$, is then calculated by discretizing
equation (\ref{eq:Lx1}),

\begin{equation} \label{eq:dLx}
\Delta L_{\rm Xj} = \frac{5k_{\rm b}}{2 \mu m_{\rm
p}}\big(\dot{M}_{\rm{clas} \rm j}\Delta T_{\rm j} +\Delta \dot{M}_{\rm
{clas} \rm j}T_{\rm j}\big).
\end{equation}

In any given shell, the term involving $\Delta T_{\rm j}$ is assumed
to be small compared to the term involving $\Delta \dot{M}_{clas \rm
j}$. By ignoring the $\Delta T_{\rm j}$ term the resulting mass
deposition rate is larger than it would be if the change in
temperature across the shell was taken into account. Therefore, the
classical value is the absolute maximum possible mass deposition
rate. The luminosity of a given shell is then assumed to be
\citep[e.g.][]{voigt04},

\begin{equation} \label{eq:dLx1}
\Delta L_{\rm Xj} = 4 \pi r^{2} \Delta r n^{2} \Lambda_{\rm rad} =
\frac{5 k_{\rm b}T_{\rm j}}{2 \mu m_{\rm p}}\Delta \dot{M}_{\rm{clas}
\rm j}.
\end{equation}

Therefore, the rate at which mass is deposited within the $jth$ shell
is

\begin{equation} \label{eq:turn1}
\Delta \dot{M}_{\rm{clas}, \rm j} = \frac{4 \pi {r_{\rm j}}^{2} \Delta
r {n_{\rm j}}^{2} \Lambda_{\rm rad}}{\frac{5 k_{\rm b}T_{j}}{2 \mu
m_{p}}}.
\end{equation}

The total X-ray luminosity emitted by the gas within a particular
radius is calculated by summing the contributions from all of the
shells within this radius

\begin{equation} \label{eq:Lxclas}
L_{\rm X}(<r) = \frac{5k_{\rm b}}{2 \mu m_{\rm p}} \dot{M}_{\rm
clas}(<r)T(r),
\end{equation}
where $\dot{M}_{\rm clas}(<r)$ is the integrated mass deposition rate
obtained by summing the contribution from each shell within the
radius, $r$.

If we include a heating term the expression equivalent to equation
(\ref{eq:Lx1}) becomes,

\begin{equation} \label{eq:Lx2}
L_{\rm X}(<r)- H(<r) = \frac{5k_{\rm b}}{2 \mu m_{\rm p}}\dot{M}_{\rm
cool} T(r),
\end{equation}
where $\dot{M}_{\rm cool}$ is the integrated mass deposition rate,
resulting from the excess of cooling over heating. This integrated
mass deposition rate is distinct from the integrated classical mass
deposition rate which is only used to describe the situation in the
absence of heating.

With equation (\ref{eq:dLx}) becomes,
\begin{equation} \label{eq:Lx10}
\Delta\big(L_{\rm X,j}- H_{\rm j}\big) = \frac{5k_{\rm b}}{2 \mu m_{\rm
p}}\big(\dot{M}_{\rm{cool},\rm j}\Delta T_{\rm j} +\Delta
\dot{M}_{\rm{cool},\rm j}T_{\rm j}\big).
\end{equation}

In practice, a realistic estimate of the mass deposited into each
shell, which takes into account heating, can be obtained by fitting
models to the X-ray spectrum of each shell or a larger region if
desired. The spectroscopic mass deposition rate is just a fitting
parameter in this case and an integrated value can be obtained in
exactly the same way as for the classical case above.

Generalising equation(\ref{eq:Lxclas}) to allow for heating we find
that if the spectroscopically determined mass deposition rate,
$\dot{M}_{\rm obs}$, is representative of the actual mass deposition
rate, $\dot{M}_{\rm cool}$, then one can determine the heating rate
from,

\begin{equation} \label{eq:Lxclas1}
L_{\rm X}(<r)- H(<r) = \frac{5k_{\rm b}}{2 \mu m_{\rm p}} \dot{M}_{\rm
obs}(<r)T(r)= \frac{5k_{\rm b}}{2 \mu m_{\rm p}} \dot{M}_{\rm
cool}T(r).
\end{equation}

If the spectroscopically determined value for the mass deposition rate
is a true representative of the real mass deposition rate, then the
integral within the cooling radius is simply the integrated mass
deposition rate within the cooling radius, irrespective of the radial
distribution. Therefore, it would seem a sensible starting point for
any calculations of the heating rates in galaxy clusters. To ensure
consistency between theory and observations, we must employ the same
mass flow rate (integrated mass deposition rate) over the same region,
in our heating models. Consequently, if we use the observed
spectroscopic mass deposition rate integrated up to the cooling
radius, then to be strictly accurate, the heating rates we
subsequently calculate are only valid up to this radius.

\subsection{Model Integrated Mass Deposition Rates}

We now construct a model for the integrated mass deposition rate in
galaxy clusters on the basis of the observational evidence. Although
the model function is essentially empirical there are three main
constraints to which it must adhere. These are: firstly, the fact that
the observed spectroscopic mass deposition rate in each shell is
approximately constant up to the cooling radius suggests linear
dependence of flow rate on radius. Secondly, the observed density
profiles also suggest that the mass deposition rate is negligible near
the cluster centre. Otherwise a pronounced peak would be observed in
the density profiles. Thirdly, the integrated mass deposition rate at
the cooling radius must be equal to the observed integrated mass
deposition rate at that location. A simple empirical form for the
integrated mass deposition rate which is, in some way, consistent with
the above is

\begin{equation} \label{eq:mdotofrad}
\dot{M}(r) = \dot{M}_{\rm obs}(<r_{\rm cool})\bigg[K +
\bigg(\frac{r}{r_{\rm cool}}\bigg)^{2}\bigg]^{0.5},
\end{equation}
where $\dot{M}_{\rm obs}(<r_{\rm cool})$ is the integrated
spectroscopic mass deposition rate at the cooling radius, $K$ is a
constant and $r_{\rm cool}$ is the cooling radius.

To estimate $K$ we express it the following way, $K = (r_{\rm
K}/r_{\rm cool})^{2}$, where $r_{\rm K}$ is a length scale
corresponding to $K$. A suitable value for $r_{\rm K}$ is the radius
of the central galaxy, roughly 30 kpc, since the effect of the
interstellar medium is likely to significantly alter the inflow of
material. This leads to values for $K$ ranging from 0.73, for Virgo,
to 0.04, for A478.

Note that our expression for $\dot{M}$ implies that ${\rm d}\dot{M} /
{\rm d} r =0$ for samll radii, meaning that no material is
deposited. However, at or near $r=0$ the material must be deposited
since it cannot flow to negative radii. This can account for the
excess star formation which is observed to occur in these regions. For
reasonable gradients of the temperature distribution, the second term
on the right hand side of equation (\ref{eq:h0}) then tends to
infinity as $r$ tends to zero. This implies that our model predicts
negative heating rates, $h$, for small radii. In otherwords, the model
is only valid outside the region at the very centres of clusters. To
indicate the range over which our model can provide physically
meaningful predictions, we list in Table 3, the minimum radius for
each cluster at which $h$=0. We take this minimum radius as the lower
integration limit in our calculation of the total heating rates.

For each cluster the observations are inevitably limited by spatial
resolution and in some cases the lower integration limits, explained
above, occur at radii which are closer to the cluster centre than the
spatial resolution limits. In principle, the spatial resolution
presents a limit beyond which we are unable to accurately describe the
temperature and density of the cluster gas. However, because we have
fitted analytical functions to the temperature and density data for
each cluster we can extrapolate the behaviour of the temperature and
density, and hence the heating rates, from the data point nearest the
cluster centre further towards the centre. The results at $r < r_{\rm
min}$ should be ignored as we have no observational information to
constrain them.

\subsection{Thermal conduction}

Thermal conduction of heat from the cluster outskirts to their centres
may provide the required heating of the central regions without an
additional energy source, like an AGN \citep[e.g.][]{gaetz,zakamska03,
voigt04}. Of course, thermal conduction only transports energy from
one region to another and does not lead to net heating. However, if we
consider the case for which there is an infinite heat bath at large
radii, then any drop in temperature at large radii, due to the inward
transfer of thermal energy, is negligible.

Several 1-d models have been able to achieve a steady-state using
thermal conduction alone \citep[e.g.][]{zakamska03} and the combined
effects of thermal conduction and AGNs
\citep[e.g.][]{brueggen03,ruszbegel02}. We, therefore, compare the
required heating rates with an estimate of the energy transport
supplied by thermal conduction given the current state of each galaxy
cluster. 

The volume heating rate for thermal conduction from an infinite heat
bath is given by

\begin{equation} \label{eq:cond}
\epsilon_{cond} = \frac{1}{r^{2}}\frac{\rm d}{{\rm
d}r}\bigg(r^{2}\kappa \frac{{\rm d}T}{{\rm d}r}\bigg),
\end{equation}
where $\kappa$ is the thermal conductivity and ${\rm d}T/{\rm d}r$ is
the temperature gradient.

We take the thermal conductivity to be given by \cite{spitzer}, but
include a suppression factor, $f(r)$, defined in the introduction, to
take into account the effect of magnetic fields,

\begin{equation} \label{eq:spitz}
\kappa = \frac{1.84 \times 10^{-5} f(r) T^{5/2}}{\ln \Lambda_{\rm c}},
\end{equation}
where $\ln \Lambda_{\rm c} $ is the Coulomb logarithm.

For pure hydrogen the Coulomb logarithm is \citep[e.g.][]{plasmas}

\begin{equation} \label{eq:coul}
\Lambda_{\rm c} = 24 \pi n_{e} \bigg(\frac{8\pi e^{2}
n_{\rm e}}{k_{\rm b}T}\bigg)^{-3/2},
\end{equation}
where $n_{\rm e}$ is the electron number density.

For a steady-state to exist, the heating by thermal conduction must be
equal to the heating rate,

\begin{equation} \label{eq:condeq}
\epsilon_{cond} = -h,
\end{equation}
where $h$ is the required heating rate per unit volume, given by
equation (\ref{eq:h0}).

This is essentially the same energy equation as solved by
\citep[e.g.][]{zakamska03}, although we also allow for mass
deposition, line-cooling and variations in the Coulomb logarithm. We
then solve equation (\ref{eq:condeq}), using the observed temperature
and density profiles, to determine the radial dependence of the
suppression factor. This is different to the approach of
\citep[][]{zakamska03}. In their paper they use the same equations to
calculate temperature and density profiles that are consistent with a
constant suppression factor. These derived profiles are only
constrained by observations at the minimum and maximum radii
accessible to observations. We take the density and temperature
profiles from the observed data and allow the suppression factor to
vary with radius.

Our function for the suppression factor is obtained by integrating
both sides of equation (\ref{eq:condeq}) over a spherical surface and
then rearranging for $f$,

\begin{equation}\label{eq:f}
f(r) = \frac {\int_{r_{\rm min}}^{r}r^{2}h(r){{\rm d}r}}{1.84 \times
10^{-5}r^{2}{{\rm d}T}/{{\rm d}r} T^{5/2}/\ln \Lambda_{\rm c}}.
\end{equation}

We note that the energy flux due to thermal conduction, $\kappa {\rm
d}T/{\rm d}r$, increases with radius so that, for an infinite heat
bath at large radii, energy must always be deposited at each smaller
radius, rather than taken away.

\section{Results:1}

\subsection{Comparison of Luminosities}

To ensure that our estimates are compatible with those calculated from
observations we compare the bolometric X-ray luminosities, within the
cooling radius, determined from our functions fitted to the data and
the observed luminosities presented in \cite{birzan}. We assume half
solar metallicity for all clusters and use the cooling function as
given in \cite{sutherland}.

\begin{table*} 
\centering
\begin{minipage}{140mm}
\begin{tabular}{rcccc}
\hline Name & $L_{X}(r_{\rm cool})/10^{42} {\rm erg s^{-1}}$ & $L_{X}(r_{\rm cool})/10^{42} {\rm erg s^{-1}}$ & $r_{\rm cool}/kpc$ & $r_{\rm min}/kpc$\\
\hline
Virgo & $10.7$ & $9.8^{+0.8}_{-0.7}$ & 35 & 1\\ 
\\
Perseus &$550$ & $670^{+40}_{-30}$ & 102 & 3\
\\ 
Hydra & $243$ &$250^{+15}_{-15}$ & 100 & 3\\ 
\\
A2597 & $470$ & $430^{+40}_{-30}$ & 129 & 4\\ 
\\
A2199 & $156$ & $150^{+10}_{-10}$ & 113 & 1\\ 
\\
A1795 & $493$ & $490^{+30}_{-30}$ & 137 & 1\\ 
\\
A478 & $1494$ & $1220^{+60}_{-60}$ & 150 & 3\\ 
\hline
\end{tabular}
\caption{Comparison of bolometric luminosities derived from our
functions fitted to the data (column 2) with those given in B{\^i}rzan
et al. (2004) (column 3), the cooling radius (column 4) and the lower
integration limits (column 5).}\label{tab:3}
\end{minipage}
\end{table*}

Table \ref{tab:3} demonstrates reasonable agreement between the
results obtained using the fitted functions and those presented in
\cite{birzan}.

\subsection{Integrated Mass Deposition rates}

Since the values of the integrated mass deposition rates that we use
are critical to determining the heating rates, we present these
first. In table 4 we compare the integrated spectroscopic mass
deposition rate, $\dot{M}_{\rm spec}(<r_{\rm cool})$
\citep[][]{birzan}, with the integrated classical mass deposition
rate, $\dot{M}_{\rm clas}(<r_{\rm cool})$ \citep[][]{fab94} at the
cooling radius, and the central integrated mass deposition rate
predicted by equation (\ref{eq:mdotofrad}).

\begin{table*}
\centering
\begin{minipage}{140mm}
\begin{tabular}{rccc}
\hline Name & $\dot{M}_{\rm spec}(<r_{\rm cool})$ & $\dot{M}_{\rm
clas}(<r_{\rm cool})$ & $\dot{M}_{\rm spec}(<r_{\rm cool})\frac{r_{\rm
K}}{r_{\rm cool}}$ \\ \hline Virgo & $1.8^{+1.2}_{-0.16}$ & 10 & 1.3\\
\\ Perseus&$54^{+48.}_{-18}$ & 183 & 4.7\\ \\ Hydra & $14^{+9}_{-7}$ &
315 & 1.3\\ \\ A2597 &$59^{+40}_{-40}$ & 480 & 3.2\\ \\ A2199 &
$2^{+7}_{-1.9}$ &150 & 0.1\\ \\ A1795 & $18^{+12}_{-10}$& 478 & 0.9\\
\\ A478 & $150^{+60}_{-68}$& 570 & 6\\ \hline

\end{tabular}
\caption{Mass deposition rate parameters (all quantities in solar
masses per year). The integrated spectral mass deposition rates are
taken from B{\^i}rzan et al. (2004) and the classically determined
values are from Fabian (1994) and references therein. The central mass
flow rates calculated using equation (\ref{eq:mdotofrad}) are also
shown for comparison.}\label{tab:4}
\end{minipage}
\end{table*}

From equation (\ref{eq:Lx2}) it is clear that the spectroscopic mass
deposition rates are indicative of the energy injected into the
cluster. Because of this, the spectroscopic mass deposition rate
should be less than the classical value,depending upon the relative
magnitudes of the heating and cooling, as is the case. However,
because of the time taken to dissipate this injected energy into the
ambient gas this implies that the current spectroscopic mass
deposition rates are a function of the heating which has taken place
over the last few times $10^{8}$ yrs.

The central integrated mass deposition rates predicted by equation
(\ref{eq:mdotofrad}) are sufficiently low that the total mass
deposited in the cluster centre, over a Gyr, will not significantly
alter the total central mass. It should be noted that a smaller value
of the constant, $K$, in equation (\ref{eq:mdotofrad}), would result
in still smaller central integrated mass deposition rates.

\subsection{Required Heating rates and Thermal Conduction Suppression factors}

\begin{figure*}
\centering \includegraphics[width=14cm]{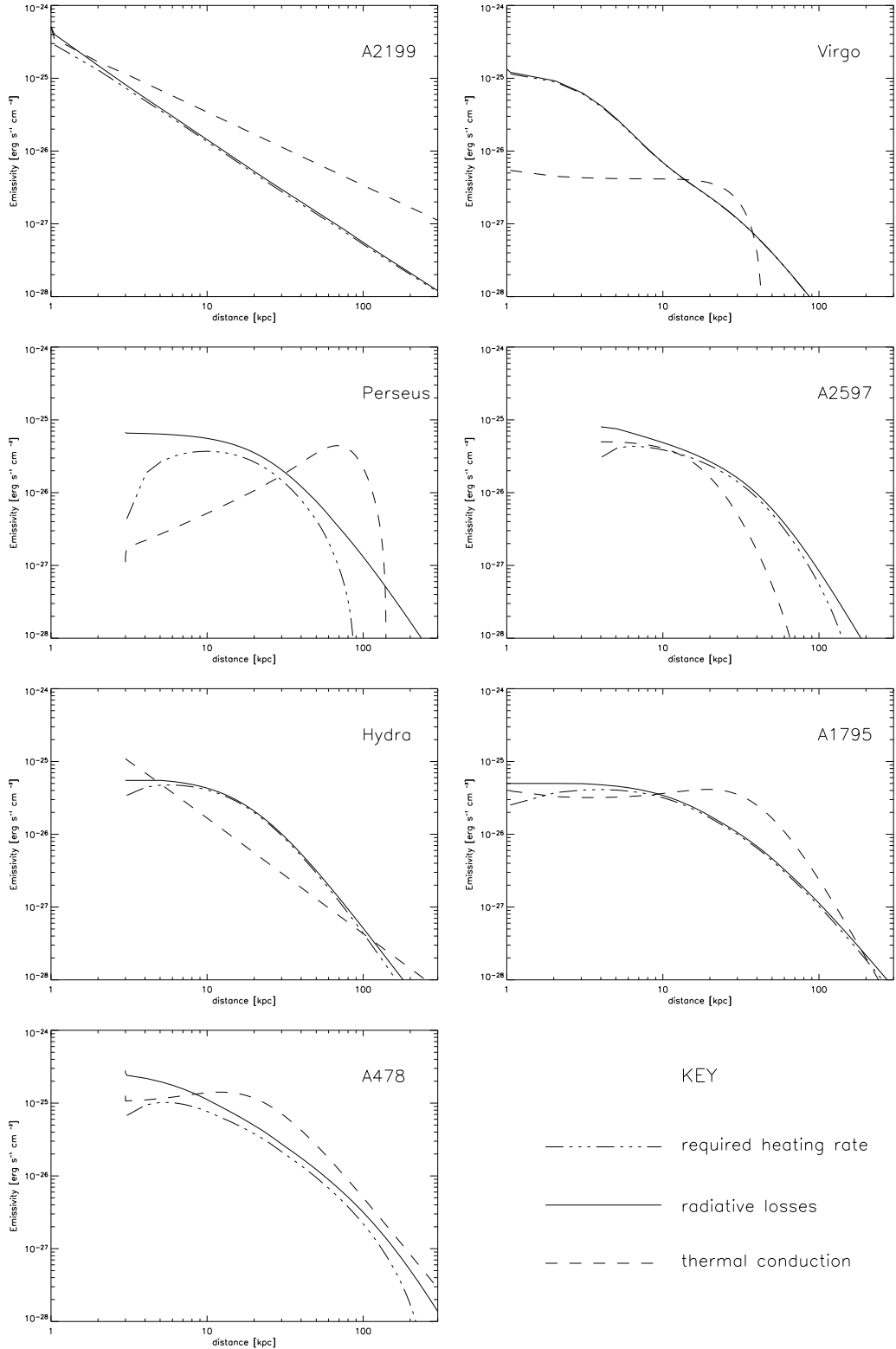}
\caption{Comparison of cooling and heating rate profiles for each
cluster. The heating rates fall to zero at a radius which depends upon
the mass deposition rate. At the point at which this occurs, we also
terminate each of the other curves since we cannot determine the gas
properties within this radius.}
\label{fig:heatmdot}
\end{figure*}

The heating rates are shown in figure \ref{fig:heatmdot}, for the form
of integrated mass deposition rates given in equation
(\ref{eq:mdotofrad}). For comparison, we also show the volume heating
rate profiles for unsuppressed thermal conduction.

Figure \ref{fig:heatmdot} shows that the heating rates for the entire
sample of clusters exhibit similar profiles. This is because the
radiative losses most strongly depend upon the density, therefore so
must the required heating rate. Since the majority of the density
profiles are described by $\beta$-profiles, this similarity is
expected.

In comparison, it is clear that the volume heating rates for
unsuppressed thermal conduction vary from cluster to cluster and do
not share the same profile as the required heating rates. Even for the
clusters in which thermal conduction can supply the necessary energy,
the different radial dependences require that the suppression factor
is fine-tuned to provide the correct heating rate for all radii. This
demonstrates the different nature of the physical processes involved
in radiative cooling and thermal conduction, making a balance between
the two hard to achieve.

\begin{figure*}
\centering \includegraphics[width=7cm]{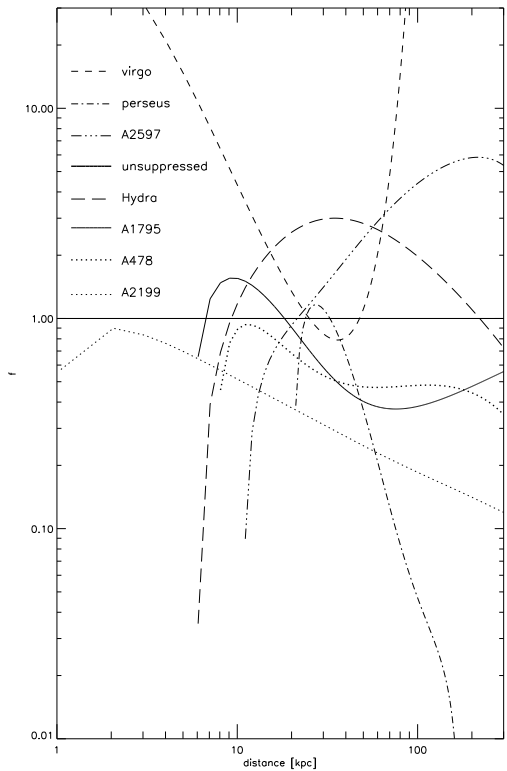}
\caption{Thermal conduction suppression factors for the sample of
seven clusters. The thick, solid line shows the maximum physically
meaningful suppression factor (f=1) for comparison.}
\label{fig:fofr}
\end{figure*}

Using equation (\ref{eq:f}) we calculate the thermal conduction
suppression factor, as a function of radius, for which the required
heating rate is achieved. These results, plotted in figure
\ref{fig:fofr}, suggest that thermal conduction could provide the
required heating rates in A2199, A478, Perseus and very nearly
A1795. It seems impossible that thermal conduction would be able to
transport sufficient energy towards the central regions for the
remainder of the sample.

It is evident that the suppression factor profiles are different in
almost every case, although A1795 and A478 appear to have certain
features in common. The only commen trait is that the required
suppression factor tends to be largest near the centre, except for
A2597 and Hydra. This suggests that even in the regions where the
suppression factor takes physically meaningful values ($f<1$), in
order for thermal conduction to provide the necessary heating, the
parameters which determine the suppression factor require unique fine
tuning in each cluster. We note that suppression factors of greater
than unity are unphysical for laminar flows but are possible in mixing
layers \citep[e.g.][]{cho03}.

Our suppression factors can be compared with those found by
\cite{zakamska03} who studied four of the clusters in our sample. They
find suppression factors or 1.5, 2.4, 0.4 and 0.2 for Hydra, A2597,
A2199 and A1795 respectively. These values agree with our results in
that we also find that thermal conduction must take unphysically large
values for Hydra and A2597 and realistic value for the remaining two
clusters. \cite{voigt04} also provide a suppression factor, roughly
0.2-0.3, for A478 which is consistent with, albeit lower than,the
possible values we have found for the same cluster.

\section{Comparison of Required Heating Rates with Observations}
\subsection{Comparison with AGN}
The time-averaged mechanical luminosity of a central AGN in each
cluster is estimated using

\begin{equation} \label{eq:mechave}
\langle L_{\rm mech}\rangle = \frac{\alpha PV}{t},
\end{equation}
where $\alpha$=16. $P$ is the ambient cluster gas pressure, $V$ is the
bubble volume and $t$ is the estimated duty cycle for an AGN. We
assume initially that the duty cycle of each AGN is $10^{8}$ yrs. The
factor 16 arises from 2 bubbles per outburst, a factor of 2 for the
energy dissipated in the shock expansion of the bubble and 4 for
$\gamma/(\gamma -1)$ where $\gamma$ = 4/3 for relativistic gas.

This is simply an estimate of the rate at which energy is injected
into the cluster, by the central AGN, and is not related to any
particular physical process by which this energy is dissipated
e.g. the viscous dissipation of sound waves or effervescent heating.

The estimated AGN power injection rates are given in column 2 of Table
5. To compare the required heating rates with the observational
estimates of AGN heating in equation (\ref{eq:mechave}) we must
calculate the volume integral of equation (\ref{eq:h0}). From this we
can estimate the heating luminosity required within a given volume to
satisfy the requirements of the assumed steady-state. Comparing this
value with that of the power output from AGNs at the centres of galaxy
clusters allows us to estimate the frequency with which buoyant plasma
bubbles, identical to those currently observed, must be produced in
order to provide sufficient heating within the cooling radius. This is
defined as the AGN duty cycle, which may be calculated using

\begin{equation} \label{eq:comp}
 \frac{4\pi \int_{r_{\rm min}}^{r_{\rm cool}}r^{2} h {\rm d}r}{\langle
 L_{\rm mech}\rangle} = \frac{t}{\tau},
\end{equation}
where $r_{\rm cool}$, the upper integration limit, is the cooling
radius and $r_{\rm min}$ is the radius at which the heating rate tends
to zero for our chosen descriptions of the mass flow rate, $\tau$ is
the calculated duty cycle and $t$ is the time defined in equation
(\ref{eq:mechave}).

\subsection{The Combined Heating by Thermal Conduction and AGN}

To estimate the total rate of energy injection by AGNs into the
cluster gas within the cooling radius we assume that all of the energy
available in the bubbles is dissipated within the cooling
radius. Furthermore, the integral of equation (\ref{eq:cond}) over a
spherical surface gives the rate at which thermal conduction transfers
heat across a radius, $r$,

\begin{equation} \label{eq:Lcond}
L_{\rm cond}(r) = 4 \pi r^{2} \kappa \frac{{\rm d}T}{{\rm d}r}.
\end{equation}

The maximum total rate at which energy can be injected into this
region is the sum of these two processes. We compare this sum with the
heating rate required for steady-state and allowing for mass
deposition. Within the cooling radius, we can then determine the
thermal conduction suppression factor required to maintain a
steady-state given the current AGN heating rate. This suppression
factor is calculated using

\begin{equation} \label{eq:Lcondf}
f = \frac{L_{\rm heat}(<r) - \langle L_{\rm mech}\rangle}{L_{\rm
cond}(r)},
\end{equation}
where $L_{\rm heat}(<r)$ is the required heating rate for the
spherical volume within $r$, $\langle L_{\rm mech}\rangle $ is the
current mechanical power of the AGN assuming a duty cycle of
$10^{8}$yrs.

\section{Results:2}
\subsection{AGN Duty Cycles}

The duty cycles (see Table 5) for Virgo and A478 are of the order of
$10^{6}$ yrs which is very short compared to the predicted lifetimes
of AGN \citep[e.g.][and references therein]{nipbin}.  In contrast, the
Hydra cluster requires recurrent outbursts of magnitude similar to the
currently observed one only every $10^{8}$ yrs, or so. The required
duty cycles for the remaining AGNs are of the order of $10^{7}$
yrs. From this we conclude that if thermal conduction is negligible
and if this sample is representative of galaxy clusters in general,
then many if not all clusters will probably be heated, at certain
points in time, by extremely powerful AGN outbursts. However, it is
worthwhile pointing out that roughly 71\% of cD galaxies at the
centres of clusters are radio-loud \citep[][]{burns90} which is larger
than for galaxies not at the centres of clusters. This may suggest
that the galaxies at the centres of clusters are indeed active more
frequently than other galaxies. It is possible that a combination of
these two effects satisfies the heating requirements we have
identified.

In contrast to the spectroscopically derived mass deposition rates,
the plasma bubbles, that still exhibit radio emission and are
currently observable, are probably more indicative of the current
heating rate rather than the heating rate over the previous $10^{9}$
years, or so. If this is true, only Hydra is currently being
excessively heated in our small sample.

Our results also show that the gas outside the cooling radius requires
significant amounts of energy to allow a steady-state. Given the large
heating requirements for the regions outside the cooling radii of
Virgo, A2199 and A478, it is reasonable to ask whether it is possible
for the observed plasma bubbles in these clusters to deliver energy at
such large radii. The observational evidence \citep[e.g.][]{birzan}
seems to show that bubbles rarely rise to beyond 40 kpc from the
cluster centre and in general tend to reach only a few tens of
kiloparsecs. This may be because beyond these radii, hydrodynamic
instabilities have shredded the bubbles \citep[e.g.][]{instab105,
reynolds05, robinson04, bub01}, or alternatively the isoentropy radius
at which the bubbles spread out and form pancakes is generally at
radii of these magnitudes \citep[e.g.][]{bub01}. Alternatively, the
apparent absence of bubbles at larger radii may be a selection effect
since they are harder to detect against the rapidly declining X-ray
surface brightness profiles. In any case, it is possible that
extremely powerful outbursts with highly extended jets such as Hydra-A
or Cygnus-A type events are the best mechanism by which energy is
supplied to such large radii, since they cannot be reached by smaller
scale outbursts.

\subsection{The Combined Heating by Thermal Conduction and AGNs}

\begin{table*} 
\centering
\begin{minipage}{140mm}
\begin{tabular}{rccccc}
\hline Name & $L_{\rm mech}/\alpha(10^{42} {\rm erg s^{-1}})$&
 $\tau/(10^{8} {\rm yr})$ & $L_{\rm cond}/(10^{42}{\rm erg s^{-1}})$ & $L_{\rm heat}/(10^{42} {\rm erg s^{-1}})$ & $f$\\
 \hline 
Virgo &  $0.052^{+0.028}_{-0.032}$& $0.078$ & $13.7$ & $10.7$ &$0.72$\\ 
\\ 
Perseus &$5.9^{+3.8}_{-1.2}$& $0.28$ & $4360$ & $236$ & $0.05$\\ 
\\
 Hydra & $27.1^{+1.5}_{-2.9}$& $1.9$ & $112.8$ & $231$ &$-1.8$\\ 
\\
 A2597 & $5.6^{+5.2}_{-1.6}$& $0.21$ &$71.0$& $427$ & $4.8$\\ 
\\ 
A2199 & $1.3^{+1.9}_{-1.2}$& $0.14$ &$797$ & $149$& $0.16$ \\ 
\\ 
A1795 &$12.5^{+16.8}_{-1.3}$& $0.43$ &$1280$ & $465$ & $0.2$ \\ 
\\ A478 & $1.9^{+1.0}_{-0.3}$& $0.026$ &$2720$ &$1169$ & $0.4$ \\ 
\hline
\end{tabular}
\caption{Assumed average mechanical luminosity per bubble in each
cluster based on the observations summarized in B{\^i}rzan et
al. (2004) and assuming a duty cycle of $ 10^{8}$ yrs in (column 2),
duty cycle of each AGN in units of $10^{8}$ yrs evaluated at the
cooling radius (column 3), rate at which energy is supplied to the
region within the cooling radius by thermal conduction (column 4), the
total required heating rate within the cooling radius (column 5) and
the suppression factor for combined heating by AGNs and thermal
conduction (column 6). All quantities are evaluated at the cooling
radius.}\label{tab:10}
\end{minipage}
\end{table*}

From table \ref{tab:10} it is clear that A2597 is currently being
heated insufficiently by the combined effects of AGN and thermal
conduction. Thus, it seems that the only way that A2597 can avoid a
cooling catastrophe is by an exceptionally large AGN outburst which
will have to be roughly 100 times more powerful than the current
outburst in order to provide sufficient heating. Of the other clusters
in this sample, the suppression factors are all physically acceptable
except for Hydra. The reason for this and the negative value of the
suppression factor for Hydra is that the energy of the current AGN
outburst exceeds the heating requirements within the cooling
radius. However, the errors associated are relatively large so it is
difficult to say with any certainty if a combination of heating by
thermal conduction and AGN can achieve a steady-state.

Although the global suppression factor for thermal conduction is
physically acceptable, it still requires fine-tuning inside each
cluster to allow a steady-state (see Section 4.3).

\section{Summary} \label{sect:summary}

In the heating of intracluster gas, the energy input required from
AGNs is likely to depend on the strength of any other heating
processes. For example, if thermal conduction is present and can
reduce the integrated mass deposition rate towards the cluster centre,
it is likely that in this case the AGN outbursts will have to be less
powerful than in an otherwise identical cluster in which thermal
conduction does not occur. There may also be additional heating
processes such as stirring by galaxy motions or the gas motion
resulting from past cluster-cluster mergers.

We have calculated the heating rates required to maintain a steady
state for a sample of seven galaxy clusters with the assumption that
the mass flow rates are independent of radius. For our model we use
the spectroscopically determined integrated mass deposition
rates. Here we summarise our main findings for each cluster in our
sample in terms of the mass deposition rates, thermal conduction and
required AGN duty cycles.

A2199: It appears that, without AGN heating, thermal conduction is
sufficient for the steady-state heating requirements. In the absence
of thermal conduction, the required AGN duty cycle of roughly $10^{7}$
yrs is short, but still compatible with predicted AGN lifetimes. When
we consider simultaneous heating by both AGN, based on the current
outburst, and thermal conduction, the suppression factor necessary to
satisfy the heating requirements within the cooling radius is roughly
0.16 which is acceptable.

Virgo: Even allowing for mass deposition, thermal conduction alone is
insufficient for providing the heating requirements at all radii. The
required AGN duty cycle, in the absence of thermal conduction, of
roughly $10^{6}$ yrs is very short compared with predicted AGN
lifetimes suggesting that significantly larger AGN outflows are
required at certain intervals to prevent a cooling catastrophe. This is
interesting since Virgo has the smallest heating requirement in our
sample. Within the cooling radius, the combined effect of the current
AGN outburst and thermal conduction can achieve the required heating
with a suppression factor of 0.7.

Perseus: It seems that in the absence of an AGN, thermal conduction
may be insufficient compared to heating requirements at roughly
roughly the central 30 kpc, but may be sufficient outside this
radius. Without thermal conduction, the required AGN duty cycle of
roughly $3 \times 10^{7}$ yrs is also compatible with predicted AGN
lifetimes. The suppression factor for the combined effect of AGN and
thermal conduction to provide the required heating is 0.05.

Hydra: Thermal conduction alone cannot provide the necessary heating
for maintaining a steady-state. The required AGN duty cycle in the
absence of thermal conduction is long compared with predicted AGN
lifetimes which reflects the powerful nature of the current
outburst. This is the most powerful central AGN outflow in this small
sample. Since the current AGN outburst in Hydra is so powerful, it
more than satisfies the heating requirements within the cooling
radius. Hence we derive a negative value for the suppression factor.

A2597: It appears that thermal conduction on its own cannot provide
sufficient heating except between roughly 4-10 kpc. Without thermal
conduction, the required AGN duty cycle is also compatible with
predicted AGN lifetimes. A2597 has the largest heating requirements in
this sample, followed by A478. Both require more than two orders of
magnitude more heating power within the cooling radius than Virgo. It
is interesting to note that even a combination of the current AGN
heating rate and full Spitzer thermal conduction provide roughly only
one fiftieth of the required heating rate within the cooling
radius. It therefore seems as if the only means by which a cooling
catastrophe can be averted in this system is by an extremely powerful
AGN outburst, or a cluster-scale merger.

A1795: On its own, thermal conduction is sufficient to match the
heating requirements throughout most of the cluster, although not near
5 kpc. The required AGN duty cycle, in the absence of thermal
conduction, of roughly $4.3\times 10^{7}$ yrs is also compatible with
predicted AGN lifetimes. For a combination of heating by AGN and
thermal conduction a suppression factor of roughly 0.2 is required to
achieve the necessary heating.

A478: Thermal conduction alone should be sufficient for matching the
heating requirements. In addition, the required AGN duty cycle in the
absence of thermal conduction is very short compared with predicted
AGN lifetimes. The central AGN outburst in this system is the least
powerful in our sample.

These results suggest that heating from AGN, with duty cycles of
$10^{8}$ yrs, alone is unable to maintain a steady-state, but may
result in a quasi steady-state if more powerful outbursts are
interspersed with the more common lower power
outbursts. Alternatively, it may be that a combination of AGN heating
and thermal conduction could provide the necessary heating, at least
within the cooling radius, in all but one cluster of our sample:
A2597. However, although thermal conduction is a possible heating
mechanism of the central regions of galaxy clusters it is still not
clear whether thermal conduction actually is an efficient heating
mechanism in galaxy clusters. For example, from numerical simulations
of the Virgo cluster by \cite{pope05}, taking in to account thermal
conduction with different suppression factors, the observed
temperature profiles are most consistent with a suppression factor of
between 0-0.1. This is at odds with the value of the suppression
factor we have derived for most clusters in this sample. In all cases
the effectiveness of thermal conduction must be fine-tuned as a
function of radius in order to allow a steady-state of the cluster
gas.

It is worth noting that galaxy clusters also need to be heated at
radii outside the cooling radius if a time-averaged steady-state is to
be maintained. Two possible mechanisms for this are sound waves
excited by buoyantly rises bubbles and comparatively rare, but very
powerful, AGN outbursts.

It may also be possible to predict which clusters are more likely to
experience more powerful AGN outbursts in the near future. For
example, if a cooling flow has formed, the fractional decrement in gas
temperature between the cooling radius and the cluster centre is
likely to be larger than in a cluster which has recently been heated
by a large AGN outburst. Then, if the duty cycle that we have
calculated based on the current heating rate, required to maintain a
steady state, is much less than $10^{8}$yrs, then we may predict that
a powerful outburst is iminent.

Based on the above criteria it seems that in our sample A478 and A2597
are the clusters most likely to experience a powerful AGN outburst in
the near future. Virgo appears also to be a candidate for an iminent
ouburst, however, the relatively small temperature difference between
the cluster center and the cooling radius indicates that it may
recently have experienced sufficient heating.

Our results for the required heating rates indicate that, for our
initial estimate of the AGN duty cycles, if buoyant plasma bubbles are
the only heating mechanism by which clusters are heated, then, of our
small sample, only the AGN at the centre of Hydra is currently
injecting sufficient energy to heat the gas within several multiples
of the cooling radius. For the remaining six clusters, the energy
available in the form of plasma bubbles is only sufficient to provide
the required heating within a relatively small fraction (roughly
0.1-0.5) of the cooling radius. This significant energy deficit
implies that these clusters may not be in a steady-state, or that the
AGNS are active far more frequently at the centres than outside of
clusters.

A possible scenario which is consistent with our results is that AGNs
in general produce only relatively small or moderately powerful
outbursts, such as we observe in the majority of clusters, which delay
the occurence of a cooling catastrophe by a small amount each
time. These small heating events may occur too infrequently to prevent
the formation of a cooling flow and the temperature gradient grows
with time. Eventually sufficient material is accreted by the central
galaxy to initiate an extremely energetic outburst such as those seen
in Cygnus-A and Hydra-A. This would also allow the heating of material
at radii which are hard to reach with lower power outbursts. In this
way the moderately powerful AGN outflows observed in most clusters
only delay more energetic events. Such a scenario for clusters,
small-scale AGN heating interpersed occasionally with more energetic
events, evolution is complex and would have to be modelled
numerically.

\section{Acknowledgements}

We especially thank David Rafferty and Brian McNamara for the A2597
data which comprises both the latest data and calibrations (Rafferty,
D.A, et al. 2005, in preparation), Steve Allen and Ming Sun for the
A478 data which also comprises the latest data and calibrations. Also,
Larry David for the Hydra data, Jeremy Sanders for the Perseus data,
Stefano Ettori for the A1795 data and Timothy Dennis for additional
copies of the A2597, A478, A1795 data. ECDP thanks the Southampton
Escience Centre for funding in the form of a studentship, GP and CRK
thank PPARC for rolling grant support. We also thank the anonymous
referee for many useful comments which improved the paper.

\bibliography{database} \bibliographystyle{mn2e}

\begin{thebibliography}{}

\bibitem[\protect\citeauthoryear{{Allen}, {Schmidt} \& {Fabian}}{{Allen}
  et~al.}{2001}]{allen01}
{Allen} S.~W.,  {Schmidt} R.~W.,    {Fabian} A.~C.,  2001, MNRAS, 328, L37

\bibitem[\protect\citeauthoryear{{B{\^ i}rzan}, {Rafferty}, {McNamara}, {Wise}
  \& {Nulsen}}{{B{\^ i}rzan} et~al.}{2004}]{birzan}
{B{\^ i}rzan} L.,  {Rafferty} D.~A.,  {McNamara} B.~R.,  {Wise} M.~W.,
  {Nulsen} P.~E.~J.,  2004, ApJ, 607, 800

\bibitem[\protect\citeauthoryear{{Br{\" u}ggen}}{{Br{\"
  u}ggen}}{2003}]{brueggen03}
{Br{\" u}ggen} M.,  2003, ApJ, 593, 700

\bibitem[\protect\citeauthoryear{{Br{\" u}ggen} \& {Kaiser}}{{Br{\" u}ggen} \&
  {Kaiser}}{2002}]{nature}
{Br{\" u}ggen} M.,  {Kaiser} C.~R.,  2002, Nat., 418, 301

\bibitem[\protect\citeauthoryear{{Burns}}{{Burns}}{1990}]{burns90}
{Burns} J.~O.,  1990, Bull. Am. Astron. Soc., 22, 821

\bibitem[\protect\citeauthoryear{{Carilli} \& {Taylor}}{{Carilli} \&
  {Taylor}}{2002}]{carilli02}
{Carilli} C.~L.,  {Taylor} G.~B.,  2002, ARA\&A, 40, 319

\bibitem[\protect\citeauthoryear{{Cho}, {Lazarian}, {Honein}, {Knaepen},
  {Kassinos} \& {Moin}}{{Cho} et~al.}{2003}]{cho03}
{Cho} J.,  {Lazarian} A.,  {Honein} A.,  {Knaepen} B.,  {Kassinos} S.,
  {Moin} P.,  2003, ApJ, 589, L77

\bibitem[\protect\citeauthoryear{{Choudhuri}}{{Choudhuri}}{1998}]{plasmas}
{Choudhuri} A.,  1998, The Physics of Fluids and Plasmas, an introduction for
  astrophyisicists.
Cambridge University Press

\bibitem[\protect\citeauthoryear{{Churazov}, {Br{\" u}ggen}, {Kaiser}, {B{\"
  o}hringer} \& {Forman}}{{Churazov} et~al.}{2001}]{bub01}
{Churazov} E.,  {Br{\" u}ggen} M.,  {Kaiser} C.~R.,  {B{\" o}hringer} H.,
  {Forman} W.,  2001, ApJ, 554, 261

\bibitem[\protect\citeauthoryear{{Churazov}, {Forman}, {Jones} \& {B{\"
  o}hringer}}{{Churazov} et~al.}{2003}]{churpers}
{Churazov} E.,  {Forman} W.,  {Jones} C.,    {B{\" o}hringer} H.,  2003, ApJ,
  590, 225

\bibitem[\protect\citeauthoryear{{David}, {Nulsen}, {McNamara}, {Forman},
  {Jones}, {Ponman}, {Robertson} \& {Wise}}{{David} et~al.}{2001}]{davhyd01}
{David} L.~P.,  {Nulsen} P.~E.~J.,  {McNamara} B.~R.,  {Forman} W.,  {Jones}
  C.,  {Ponman} T.,  {Robertson} B.,    {Wise} M.,  2001, ApJ, 557, 546

\bibitem[\protect\citeauthoryear{{Dennis} \& {Chandran}}{{Dennis} \&
  {Chandran}}{2005}]{denchand}
{Dennis} T.~J.,  {Chandran} B.~D.~G.,  2005, ApJ, 622, 205

\bibitem[\protect\citeauthoryear{{Edge}}{{Edge}}{2001}]{edge01}
{Edge} A.~C.,  2001, MNRAS, 328, 762

\bibitem[\protect\citeauthoryear{{Ettori}, {Fabian}, {Allen} \&
  {Johnstone}}{{Ettori} et~al.}{2002}]{ettori}
{Ettori} S.,  {Fabian} A.~C.,  {Allen} S.~W.,    {Johnstone} R.~M.,  2002,
  MNRAS, 331, 635

\bibitem[\protect\citeauthoryear{{Fabian}}{{Fabian}}{1994}]{fab94}
{Fabian} A.~C.,  1994, ARA\&A, 32, 277

\bibitem[\protect\citeauthoryear{{Fabian}, {Celotti}, {Blundell}, {Kassim} \&
  {Perley}}{{Fabian} et~al.}{2002}]{fabpers02}
{Fabian} A.~C.,  {Celotti} A.,  {Blundell} K.~M.,  {Kassim} N.~E.,    {Perley}
  R.~A.,  2002, MNRAS, 331, 369

\bibitem[\protect\citeauthoryear{{Gaetz}}{{Gaetz}}{1989}]{gaetz}
{Gaetz} T.~J.,  1989, ApJ, 345, 666

\bibitem[\protect\citeauthoryear{{Ghizzardi}, {Molendi}, {Pizzolato} \& {De
  Grandi}}{{Ghizzardi} et~al.}{2004}]{ghizzardi04}
{Ghizzardi} S.,  {Molendi} S.,  {Pizzolato} F.,    {De Grandi} S.,  2004, ApJ,
  609, 638

\bibitem[\protect\citeauthoryear{{Johnstone}, {Allen}, {Fabian} \&
  {Sanders}}{{Johnstone} et~al.}{2002}]{johnstone}
{Johnstone} R.~M.,  {Allen} S.~W.,  {Fabian} A.~C.,    {Sanders} J.~S.,  2002,
  MNRAS, 336, 299

\bibitem[\protect\citeauthoryear{{Kaiser}, {Pavlovski}, {Pope} \&
  {Fangohr}}{{Kaiser} et~al.}{2005}]{instab105}
{Kaiser} C.~R.,  {Pavlovski} G.,  {Pope} E.~C.~D.,    {Fangohr} H.,  2005,
  MNRAS, 359, 493

\bibitem[\protect\citeauthoryear{{McNamara}, {Wise}, {Nulsen}, {David},
  {Carilli}, {Sarazin}, {O'Dea}, {Houck}, {Donahue}, {Baum}, {Voit},
  {O'Connell} \& {Koekemoer}}{{McNamara} et~al.}{2001}]{mcnam01}
{McNamara} B.~R.,  {Wise} M.~W.,  {Nulsen} P.~E.~J.,  {David} L.~P.,  {Carilli}
  C.~L.,  {Sarazin} C.~L.,  {O'Dea} C.~P.,  {Houck} J.,  {Donahue} M.,  {Baum}
  S.,  {Voit} M.,  {O'Connell} R.~W.,    {Koekemoer} A.,  2001, ApJ, 562, L149

\bibitem[\protect\citeauthoryear{{Nipoti} \& {Binney}}{{Nipoti} \&
  {Binney}}{2005}]{nipbin}
{Nipoti} C.,  {Binney} J.,  2005, MNRAS, pp 535--+

\bibitem[\protect\citeauthoryear{{Pope}, {Pavlovski}, {Kaiser} \&
  {Fangohr}}{{Pope} et~al.}{2005}]{pope05}
{Pope} E.~C.~D.,  {Pavlovski} G.,  {Kaiser} C.~R.,    {Fangohr} H.,  2005,
  MNRAS, 364, 13

\bibitem[\protect\citeauthoryear{{Reynolds}, {McKernan}, {Fabian}, {Stone} \&
  {Vernaleo}}{{Reynolds} et~al.}{2005}]{reynolds05}
{Reynolds} C.~S.,  {McKernan} B.,  {Fabian} A.~C.,  {Stone} J.~M.,
  {Vernaleo} J.~C.,  2005, MNRAS, 357, 242

\bibitem[\protect\citeauthoryear{{Robinson}, {Dursi}, {Ricker}, {Rosner},
  {Calder}, {Zingale}, {Truran}, {Linde}, {Caceres}, {Fryxell}, {Olson},
  {Riley}, {Siegel} \& {Vladimirova}}{{Robinson} et~al.}{2004}]{robinson04}
{Robinson} K.,  {Dursi} L.~J.,  {Ricker} P.~M.,  {Rosner} R.,  {Calder} A.~C.,
  {Zingale} M.,  {Truran} J.~W.,  {Linde} T.,  {Caceres} A.,  {Fryxell} B.,
  {Olson} K.,  {Riley} K.,  {Siegel} A.,    {Vladimirova} N.,  2004, ApJ, 601,
  621

\bibitem[\protect\citeauthoryear{{Ruszkowski} \& {Begelman}}{{Ruszkowski} \&
  {Begelman}}{2002}]{ruszbegel02}
{Ruszkowski} M.,  {Begelman} M.~C.,  2002, ApJ, 581, 223

\bibitem[\protect\citeauthoryear{{Sanders}, {Fabian}, {Allen} \&
  {Schmidt}}{{Sanders} et~al.}{2004}]{sanders}
{Sanders} J.~S.,  {Fabian} A.~C.,  {Allen} S.~W.,    {Schmidt} R.~W.,  2004,
  MNRAS, 349, 952

\bibitem[\protect\citeauthoryear{{Spitzer}}{{Spitzer}}{1962}]{spitzer}
{Spitzer} L.,  1962, Physics of Fully Ionized Gases.
Wiley-Interscience, New York

\bibitem[\protect\citeauthoryear{{Sun}, {Jones}, {Murray}, {Allen}, {Fabian} \&
  {Edge}}{{Sun} et~al.}{2003}]{sun}
{Sun} M.,  {Jones} C.,  {Murray} S.~S.,  {Allen} S.~W.,  {Fabian} A.~C.,
  {Edge} A.~C.,  2003, ApJ, 587, 619

\bibitem[\protect\citeauthoryear{{Sutherland} \& {Dopita}}{{Sutherland} \&
  {Dopita}}{1993}]{sutherland}
{Sutherland} R.,  {Dopita} M.,  1993, ApJ Supp., 88, 253

\bibitem[\protect\citeauthoryear{{Tabor} \& {Binney}}{{Tabor} \&
  {Binney}}{1993}]{tabor93}
{Tabor} G.,  {Binney} J.,  1993, MNRAS, 263, 323

\bibitem[\protect\citeauthoryear{{Voigt} \& {Fabian}}{{Voigt} \&
  {Fabian}}{2004}]{voigt04}
{Voigt} L.~M.,  {Fabian} A.~C.,  2004, MNRAS, 347, 1130

\bibitem[\protect\citeauthoryear{{Voigt}, {Schmidt}, {Fabian}, {Allen} \&
  {Johnstone}}{{Voigt} et~al.}{2002}]{voigt02}
{Voigt} L.~M.,  {Schmidt} R.~W.,  {Fabian} A.~C.,  {Allen} S.~W.,
  {Johnstone} R.~M.,  2002, MNRAS, 335, L7

\bibitem[\protect\citeauthoryear{{Zakamska} \& {Narayan}}{{Zakamska} \&
  {Narayan}}{2003}]{zakamska03}
{Zakamska} N.~L.,  {Narayan} R.,  2003, ApJ, 582, 162

\end{thebibliography}

\label{lastpage} 
\end{document}